\documentclass[journal]{IEEEtran} 
\ifCLASSINFOpdf

\else

\fi

\usepackage{xcolor}
\usepackage{soul}
\usepackage{graphics} 
\usepackage{graphicx}
\usepackage{float}
\usepackage{cleveref}
\usepackage{url}
\usepackage{amsmath} 
\usepackage{amssymb}  
\usepackage{algorithm}
\usepackage[noend]{algpseudocode}

\DeclareMathOperator*{\diag}{diag}

\usepackage{cite}
\usepackage{booktabs}
\usepackage{pgfplots}
\usepackage{pgfplotstable}
\usepackage{tikz}
\usepackage{multirow}
\usepackage{subcaption}
\pgfplotsset{compat=newest}
\usepgfplotslibrary{groupplots}
\pgfplotsset{every axis legend/.append style={legend cell align=left}}
\pgfplotsset{every axis/.append style={font=\footnotesize}}

\newcommand{\conreg}{\ensuremath{\mathbf{\Gamma}}}

\begin{document}
%

\title{Estimation and Control Using Sampling-Based Bayesian Reinforcement Learning}%

\author{Patrick~Slade,~\IEEEmembership{Student Member,~IEEE,}
        Zachary~N.~Sunberg,~\IEEEmembership{Student Member,~IEEE,}
       and~Mykel~J.~Kochenderfer,~\IEEEmembership{Senior Member,~IEEE}
\thanks{This work is supported by the National Science Foundation Graduate Research Fellowship Program Grant DGE-1656518 and the Stanford Graduate Fellowship. Toyota Research Institute (``TRI") provided funds to assist the authors with their research but this article solely reflects the opinions and conclusions of its authors and not TRI or any other Toyota entity.}%
\thanks{The authors are part of the Stanford Intelligent Systems Laboratory, Stanford University, Stanford, CA 94305 USA (e-mail: \{patslade, zsunberg, mykel\}@stanford.edu). Corresponding author: Patrick Slade.}%
}

%



\maketitle
\begin{abstract}
Real-world autonomous systems operate under uncertainty about both their pose and dynamics. Autonomous control systems must simultaneously perform estimation and control tasks to maintain robustness to changing dynamics or modeling errors. However, information gathering actions often conflict with optimal actions for reaching control objectives, requiring a trade-off between exploration and exploitation. The specific problem setting considered here is for discrete-time nonlinear systems, with process noise, input-constraints, and parameter uncertainty. This article frames this problem as a Bayes-adaptive Markov decision process and solves it online using Monte Carlo tree search with an unscented Kalman filter to account for process noise and parameter uncertainty. This method is compared with certainty equivalent model predictive control and a tree search method that approximates the QMDP solution, providing insight into when information gathering is useful. Discrete time simulations characterize performance over a range of process noise and bounds on unknown parameters. An offline optimization method is used to select the Monte Carlo tree search parameters without hand-tuning. In lieu of recursive feasibility guarantees, a probabilistic bounding heuristic is offered that increases the probability of keeping the state within a desired region. 
\end{abstract}


%
\IEEEpeerreviewmaketitle


\section{Introduction}

\IEEEPARstart{P}{lanning} for tasks such as localization and manipulation requires an accurate model of the system dynamics \cite{lavalle2001randomized,janson2015fast,thrun2005probabilistic}. However, the dynamics are often only partially known. For example, order-fulfillment robots move containers with varying loads in warehouses \cite{d2008future}, autonomous vehicles encounter changing environments \cite{beiker2012legal, sadigh2016gathering, sunberg2017internal}, and nursing robotic systems interact with people \cite{pineau2003towards}. Payload shifts, environment conditions, and human decisions act as time-varying parameters that dramatically alter system dynamics. Even for motion determined by physical laws, there are often unknown parameters, such as friction and inertial properties. In these cases, the robot must estimate the system dynamics from measurements to achieve its goals. This estimation task often conflicts with the original goal task. The robot must balance \emph{exploration} to gain better understanding of the dynamics and \emph{exploitation} of its current knowledge to obtain rewards.

There exist many principled approaches to handle the exploration-exploitation trade-off or ``dual control'' problem \cite{feldbaum1960dual}. When the state, action, and belief spaces are continuous, the exact solution is generally intractable. Many approximate solutions are used, such as adaptive control, sliding mode control, and stochastic optimal control \cite{filatov2000survey,ioannou1996robust,narendra2012stable}. Adaptive controllers typically first perform an estimation task and then perform the control task. This may be suboptimal, as the agent could use the estimation actions to begin the control task. A popular solution is certainty equivalent control, where a robot plans assuming an exact dynamics model \cite{bertsekas1995dynamic}. The approach of model predictive control (MPC) is to update the dynamics model after every observation and compute a new plan to a fixed horizon that is optimal for the updated most likely model \cite{bertsekas1995dynamic,garcia1989model}. Variants of MPC extend to nonlinear systems \cite{chen1997quasi,liu2014nonlinear} and account for uncertainty by propagating worst-case outcomes \cite{campo1987robust,allwright1992linear,kouvaritakis2015model} or using probabilistic constraints \cite{mesbah2017stochastic}. Equations from the unscented transform have been incorporated as MPC constraints in an attempt to improve state estimation \cite{streif2014stochastic,bradford2017stochastic,bavdekar2016stochastic}. However, these methods cannot guarantee stability for systems with time-varying parameters following a Gaussian distribution as the possible changes to the system are unbounded and cannot be corrected by an input-constrained control law \cite{feng1991stability, streif2014stochastic}.

Another popular approach is reinforcement learning, where the underlying planning problem is a Markov decision process (MDP) with unknown transition probability distributions \cite{Wiering2012}. Agents interact with the environment to accrue rewards and may learn the transition probabilities if it helps with the task. If the prior distribution of these transition probabilities is known, the policy that will collect the most reward in expectation is found by solving a Bayes-adaptive MDP \cite{kochenderfer2015decision}. 
Bayes-adaptive MDPs are typically computationally intractable \cite{kochenderfer2015decision}.
However, many approximate solution methods are available, particularly when the problem is recast as a partially observable MDP (POMDP).
If the problem has discrete state and action spaces, both offline \cite{bai2013planning} and online \cite{guez2013scalable} POMDP methods have been adapted to RL, and others \cite{chen2016pomdplite} can be easily adapted.
Several methods for continuous-state problems have also been proposed.
For systems where the uncertainty is approximately Gaussian, a value-iteration based method that uses an augmented reward function explicitly penalizing uncertainty has been demonstrated \cite{webb2014online}.
In a preliminary version of this work, Monte Carlo tree search (MCTS) was used to solve an approximation of the problem with the tree search, implicitly balancing the exploration and exploitation that outperformed MPC in some nonlinear system simulations \cite{slade2017simultaneous}. MCTS has also been extended to handle POMDPs with non-Gaussian noise \cite{sunberg2017pomcpow}.

The contribution of this research is to applying MCTS to the dual control problem, testing it against a baseline MPC approach, and proposing and demonstrating several approximations and techniques to improve performance.
Specifically, we address problems expressed in discrete time with input constraints; time-varying parameters; continuous state, action, and observation spaces; and process noise with a Gaussian distribution truncated to prevent non-physical parameter values.
The continuous state and action spaces of the problem are handled using the double progressive widening variant of MCTS~\cite{couetoux2011continuous}.

This paper builds upon preliminary work~\cite{slade2017simultaneous} by exploring the performance for various boundaries on parameter values, upgrading from an extended Kalman filter to an unscented Kalman filter, and applying an offline cross-entropy method~\cite{deng2006cross} to determine solver parameters without hand tuning.
It also investigates a further approximation, a tree search variant of QMDP (QMDP-TS), which is more computationally efficient, but assumes full observability of the model parameters after the first step.
Finally, a sampling-based method for regulating the state to a bounded region with high probability is described and tested.

The paper is organized as follows. \Cref{sec:background} gives an introduction to the problems and solution methods considered, \Cref{sec:problem,sec:approach} gives detailed descriptions of the problem and approach, and \Cref{sec:results} shows simulations comparing MCTS and QMDP-TS with a certainty-equivalent MPC benchmark.

\section{Background} \label{sec:background}
This section reviews sequential decision making models, approximate solution methods for these models, the cross-entropy algorithm for tuning parameters in the solvers, and gives a brief introduction to the confidence regions used for probabilistic state heuristics.

\subsection{MDPs, POMDPs, Bayes-adaptive MDPs}
	A Markov decision process is a mathematical framework for sequential decision making in which an agent will move stochastically between states over time. Various rewards are accrued for entering certain states. The agent may affect the trajectory and rewards by taking actions at each time step. An MDP is defined by the tuple $(\mathcal{X},\mathcal{U},T,R,\gamma)$, where:
\begin{itemize}
\item $\mathcal{X}$ is the set of states,
\item $\mathcal{U}$ is the set of actions the agent may take,
\item $T(x'\mid x,u)$ is the probability of transitioning to state $x'$ by taking action $u$ at state $x$,
\item $R(x,u)$ is the reward (or cost) of taking action $u$ at state $x$, and
\item $\gamma \in [0,1]$ is the discount factor for future rewards.
\end{itemize}

The solution to an MDP is a policy, $\pi(x):\mathcal{X} \rightarrow \mathcal{U}$, which maps each state to an optimal action that will accrue the most rewards in expectation over some planning horizon. For discounted or finite horizon MDPs, the optimal policy satisfies the Bellman equation \cite{bertsekas1995dynamic}. For small, discrete state and action spaces, the unique fixed point solution of the Bellman equation may be found efficiently with dynamic programming. Approximate dynamic programming may be used for problems with large or continuous state and action spaces \cite{powell2011}.

A POMDP is an extension of an MDP where the agent cannot directly observe the true state, but instead receives only stochastic observations which have distributions conditioned on the state \cite{kochenderfer2015decision}. The agent forms a belief state $b$, which encodes the probability of being in each state $x$. The agent updates its belief at each step depending on its previous action and observation. Since the agent may hold any combination of beliefs about its location in the state space, the belief space, $\mathcal{B}$, usually has infinite cardinality, making POMDPs computationally intensive to solve \cite{papadimitriou1987complexity}.

A Bayes-adaptive MDP has transition probabilities that are only partially known. Initially, the decision-making agent only knows a prior distribution of the transition probabilities. As the agent interacts with the environment, it extracts information about the transition probabilities from the history of states and actions it has visited. A Bayes-adaptive MDP becomes a POMDP by augmenting the state with the unknown parameters defining the transition probabilities.

A POMDP is actually an MDP where the state space is the belief space of the original POMDP \cite{kaelbling1998planning}, sometimes called a belief MDP. The transition dynamics of the belief MDP are defined by a Bayesian update of the belief when an action is taken and an observation received. Not all belief MDPs are POMDPs. In a POMDP, the reward for a belief is the expectation of the state reward given that the state is distributed according to that belief \cite{araya2010pomdp}. Thus uncertainty itself cannot be explicitly penalized. When the belief update is computationally tractable, approximate dynamic programming techniques designed for MDPs may be applied to POMDPs by using the corresponding belief MDP. 

\subsection{Monte Carlo Tree Search and QMDP Tree Search}

MCTS is a sampling-based online approach for approximately solving MDPs which can be applied to POMDPs by using the associated belief MDP. MCTS uses a generative model $G$ to generate a random state and reward $(x',r) = G(x,u,w)$, where $w$ is a random noise variable.
It performs a forward search through the state space, using $G$ to draw prospective trajectories and rewards.
In MCTS, a tree is created with alternating layers of state nodes and action nodes \cite{browne2012survey}.
A single iteration of MCTS consists of four stages: selection, expansion, rollout, and propagation \cite{bertsimas2014comparison}.
By performing many iterations of this process, MCTS estimates the value at each node and chooses the action with the highest value.

QMDP is an offline approximation technique that accounts for one step of uncertainty \cite{littman1995complexity}.
This method performs well when the action choice can not reduce the state uncertainty and thus information gathering is not important \cite{kochenderfer2015decision,sunberg2017pomcpow}.
The method is modified for online use by following same tree search structure as MCTS, referred to as QMDP-TS.
The first step is solved exactly the same as in MCTS on the belief MDP, but all subsequent steps are treated as fully observable.
Rather than computing beliefs for these steps, the mean is taken to be the true state and propagated exactly.

\subsection{Unscented Kalman Filter}
	In problems with linear Gaussian dynamics and observation functions, perfect Bayesian state estimation can be achieved with the Kalman filter. A Kalman filter is an iterative algorithm that can exactly update Gaussian beliefs over the state given the action taken, the observation received, and the transition and observation models \cite{thrun2005probabilistic}.
For systems with nonlinear dynamics, the extended Kalman filter approximates the state distribution with a Gaussian distribution. It propagates the prediction analytically using a linearization of the system dynamics \cite{wan2000unscented}.
The true posterior mean and covariance of the transformed Gaussian distribution can accumulate large errors and possibly diverge depending on how well this linearization matches the change in the system.
A UKF uses deterministic sampling to approximate the Gaussian distribution with a minimal set of carefully chosen points, achieving a second-order approximation (Taylor series expansion) when the points are propagated through any nonlinearity \cite{julier1997new}. 

Let $x_k$, $u_k$, and $o_k$ be the state, action, and observation at time $t=k$. For a system with nonlinear-Gaussian dynamics, the transition and observation models can be expressed as
\begin{equation}\label{eq:transition}
x_{k+1} = f(x_{k},u_{k}) + w_k
\end{equation}
\begin{equation}\label{eq:observation}
o_k = h(x_k,u_k) + v_k
\end{equation}
where $f$ and $h$ are nonlinear functions, and $w$ and $v$ are normally distributed independent random variables for the process and measurement noise, respectively. The Gaussian belief has an estimate for the mean 
and covariance. 
The UKF updates the belief at each timestep by taking a sample of sigma-points, approximating new mean and covariance predictions. This gives a new estimate for the state and covariance based on the action taken.
%
%
The UKF uses tunable parameters controlling the spread of sample points and knowledge of the distribution. Prior work has defined the optimal parameter values for estimating Gaussian distributions \cite{wan2000unscented}.

\subsection{Cross-Entropy}

The cross-entropy method is an optimization technique that iteratively updates a probability distribution that describes the input values that are likely to be optimal \cite{deng2006cross}.
This probability distribution belongs to a heuristically chosen family of parameterized distributions, for example this work uses the multivariate normal family parameterized by the mean and covariance matrix.
The update of the probability distribution is performed in two steps.
First, a fixed number of samples are drawn from the distribution and evaluated with respect to the optimization objective.
Second, a smaller number of elite samples with the highest objective values are selected and the parameters of the distribution are fit to these samples, usually by maximizing likelihood, yielding the probability distribution for the next generation.
This process is continued for a specified number of iterations or until meeting a convergence threshold.

\subsection{Confidence Regions} \label{sec:approxconfreg}

Confidence regions are multivariate extensions of confidence intervals that can be used to bound the value that a random variable will take with high probability \cite{beale1960confidence,draper2014applied}. A confidence region contains the value of a random variable with probability greater than or equal to the confidence level, $1-\alpha$. The specified statistical significance factor $\alpha$ is selected between $0$ and $1$, typically less than $0.1$. The confidence region will be computed for a jointly normal distribution $\mathcal{N}(\mu_d,\Sigma^d)$. The confidence region $\Gamma$ is a $p$-dimensional ellipsoid centered at the mean of the normal distribution
\begin{equation}
\label{eq:confreg}
\conreg(\mu_d, \Sigma^d, \alpha) = \\
\left\{ \mu : (\mu_d - \mu)' \Sigma^d (\mu_d - \mu) \leq \chi^2_{p}(\alpha) \right\} \text{,} 
\end{equation}
where $\chi^2$ is the cumulative distribution function of the chi-squared distribution as a function of $\alpha$ and the degrees of freedom $p$.

Computing the ellipsoid satisfying (\ref{eq:confreg}) requires rescaling the eigenvalues of the distribution's covariance, $\left[\lambda_1, \ldots, \lambda_p\right]$. The eigenvalue of the confidence region computed along the \textit{i}th axis is
\begin{equation}
\label{eq:ellipsoid_lengths}
c_i = \lambda_i \sqrt{\chi^2_{p}(\alpha)}\text. 
\end{equation}
The eigenvalues of the distribution's covariance are replaced by the diagonal matrix $\left[c_1, \ldots, c_p\right]$ to form the confidence region ellipsoid. 

\section{Problem Formulation} \label{sec:problem}

Consider a robot trying to control a system with linear-Gaussian dynamics. The transition at step $k$ is described by
\begin{equation}
\label{eq:basicModel}
x_{k+1} = f(x_k,\theta_k,u_k) + w_k
\end{equation}
where $x_k$ and $u_k$ are the state and action, $\theta_k$ is a vector of the unknown and time-varying parameters of the dynamics, $w_k \sim \mathcal{N}(0,\Sigma^w)$ is the process noise, and $f$ is a time-varying function that is linear with respect to $x_k$ and $u_k$,
\begin{equation}
f(x_k,\theta_k,u_k) = A(\theta_k)x_k + B(\theta_k)u_k\text.
\end{equation}

The observation model is described by the linear equation
\begin{equation}
o_k = h(x_k,\theta_k,u_k) + v_k
\end{equation}
with observation $o_k$, the measurement noise $v_k \sim \mathcal{N}(0,\Sigma^v)$, and a time-varying linear function $h$
\begin{equation}
h(x_k, \theta_k, u_k) = C(\theta_k)x_k+D(\theta_k)u_k\text.
\end{equation}

While $f$ and $h$ are physical equations known \textit{a priori}, the parameters $\theta_k$ are not known beforehand. They can be appended to the state vector to form a state-parameter vector or \emph{hyperstate} \cite{mesbah2017stochastic}
\begin{equation}
\xi_k = \left[ \begin{array}{c} x_k \\ \theta_k \end{array} \right]\text.
\end{equation}
Thus, the system dynamics for $\xi_k$ may be described by
\begin{equation}
\xi_{k+1} = \left[ \begin{array}{cc} A(\theta_k) & 0 \\ 0 & I \end{array} \right] \xi_k + \left[ \begin{array}{c} B(\theta_k) \\ 0 \end{array} \right] u_k + w_k
\end{equation}
where $w \sim \mathcal{N}(0,\diag(\Sigma^w,\Sigma^\theta))$ and $\Sigma^\theta$ is a parameter drift matrix. We assume the parameter drift is equivalent to the process noise, simplifying the nonlinear function to include additive process noise to the hyperstate

\begin{equation}
\label{eq:basicHyperstateModel}
\xi_{k+1} = f(\xi_k,u_k) + w_k
\text.
\end{equation}
The observation model is described by
\begin{equation}
\label{eq:obs}
o_k = \left[ \begin{array}{cc} C(\theta_k) & 0 \end{array} \right] \xi_k + D(\theta_k)u_k + v_k\text.
\end{equation}

	Using a UKF to describe the belief about the current hyperstate in the state-parameter space forms a belief MDP over all possible UKF states. This belief MDP is described by the tuple $(\mathcal{X},\mathcal{U},T,R)$, where:
\begin{itemize}
\item $\mathcal{X}$ is the space of all possible beliefs. Since the belief maintained by the UKF is Gaussian, it can be described by the mean and covariance, $b = \mathcal{N}(\hat{\xi},\Sigma^\xi)$.
\item $\mathcal{U}$ is all possible actions that the agent may take.
\item $T(b' \mid b,u)$ is a distribution over possible UKF states after a belief update. This distribution depends on the observation model. It is difficult to represent explicitly, so it is implicitly defined by the generative model, $G$.
\item $R(x,u)$ is a reward function for a given state and action. It is constructed as desired for a given control task. In our work, we approximated $R(x,u) = R(\hat{x},u)$, a linear reward for the estimated mean state and action.
\end{itemize}

The generative model for the UKF approximated belief MDP is
\begin{equation}
b_{k+1} = G(b_k, u_k) \text,
\end{equation}
with $G$ defined by the UKF update of the estimated mean and covariance with an observation sampled according to (\ref{eq:obs}). The observation is given by the measurement update equations. Solving this belief MDP gives a policy that approximately maximizes the sum of expected rewards over some planning horizon.

\section{Approach} \label{sec:approach}

This section discusses using MCTS, QMDP-TS, and MPC with a UKF to control a system with unknown parameters.

\subsection{Monte Carlo Tree Search and QMDP-TS}
Our approach uses the upper confidence tree (UCT) \cite{browne2012survey} with double progressive widening (DPW) \cite{couetoux2011continuous} extensions of MCTS and QMDP-TS. The tree is built by repeatedly exploring the action node that maximizes an upper confidence estimate
\begin{equation}
UCB(b, u) = \tilde{Q}(b,u) + c \sqrt{\frac{\log N(b)}{N(b,u)}}\text,
\end{equation}
where $\tilde{Q}(b,u)$ is an estimate of the state-action value function from rollout simulations and tree search, $N(b,u)$ counts the times action $u$ is taken from the hyperstate belief $b$, and $c$ is an exploration constant that balances exploration and exploitation as the tree expands.

DPW defines tree growth for large or continuous state and action spaces. To avoid a shallow search, the number of children of each state-action node ($b, u$) is limited to
\begin{equation}
\label{eq:dpw}
k N(b,u)^{\delta}\text,
\end{equation}
where $k$ and $\delta$ are parameter constants tuned to control the widening of the tree. With an increase in $N(b, u)$ the number of children also grows, widening the tree. The number of actions explored at each state is controlled in the same way with an additional set of parameters.

\begin{figure}
\centering
\includegraphics[width=0.35\textwidth]{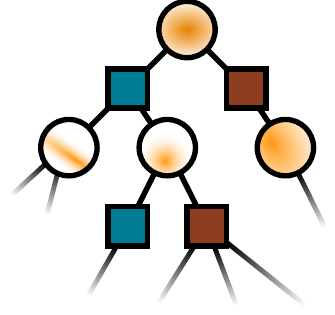}
\caption{This illustration of a simple DPW tree shows state nodes as circles and action nodes as squares. In this example, the state is a multivariate normal where the darker color and larger size of the orange gradient visually represents a larger covariance.}
\label{fig:tree_intuition}
\end{figure}

Controlling the growth of both the state and action nodes allows the tree to balance the exploration of promising actions and exploitation of the current knowledge of the system. Actions that minimize parameter uncertainty, such as the initial blue action node in Fig. \ref{fig:tree_intuition}, may result in higher rewards for subsequent children nodes. The tree search will steer exploration in these most promising nodes to find the best policy. Thus, exploratory actions are selected when they improve the reward more than exploitive actions such as the initial red action node. 

\subsection{Model Predictive Control}

MPC is a technique for online calculation of a policy \cite{garcia1989model}. A major factor contributing to its extensive use in control system design is its ability to explicitly meet state and control constraints. For our implementation of MPC, the optimization problem at each step is
\begin{equation*}
\begin{aligned}
\label{mpc}
& \text{maximize} 
& & \sum_{k = 1}^{H} R(x_k, u_k) \\
& \text{subject to}
& & x_{k+1} = A(\hat{\theta}) x_k + B(\hat{\theta}) u_k & \forall k \in 1\ldots H-1 \\
& & & \vert u_k \vert \leq u_{max} & \forall k \in 1\ldots H \text, \\
\end{aligned}
\end{equation*}
with the reward function $R$ dependent upon states and control actions taken up to the horizon of $H$ steps. The control effort has maximum bound $u_{max}$. The reward function is maximized as both the state error and control action are penalized with negative rewards without any additional terminal reward at step $H$. The dynamics matrices are a function of the unknown parameter estimates, $\hat{\theta}$, which are the current best estimates available from the UKF and remain fixed during the optimization. At each step, a series of control actions maximizing this objective function over a fixed horizon from the current state are found, and the first action from this sequence is taken \cite{bertsekas1995dynamic}.

Since there are no time-varying stochastic nonlinear MPC algorithms that guarantee feasibility for parameters with Gaussian noise, linear MPC was selected for comparison \cite{feng1991stability, streif2014stochastic}. Linear MPC acts as a baseline to benchmark the performance between the optimal control and reinforcement learning algorithms. The important distinction between the described POMDP methods and MPC is that the POMDP methods reason about learning the parameters in the system. Certainty equivalent MPC assumes values for these parameters, typically the mean of the belief from an estimator such as a UKF.

\subsection{Simultaneous Estimation and Control}

This subsection describes the high level loop that controls the system by receiving observations and specifying actions.
Any system described by a model in the form of (\ref{eq:basicHyperstateModel}) with an approximately Gaussian belief over the hyperstate can use MCTS, QMDP-TS, or MPC as the definition of a policy to choose a suitable control action.
At every time step, this policy selects an action based on the belief.
The state then evolves over time according to the dynamics and the action specified by the policy, and a new observation is generated.
This observation and action are used to update the belief state with the UKF, improving the parameter estimate and providing the belief for the next action.
The entire process is shown in Algorithm \ref{alg1}.

\begin{algorithm}
\begin{algorithmic}[1]
\caption{Simultaneous estimation and control}
\label{alg1}
\Require $b_0(x),\xi_0$
\For{$t \in [0, T)$}
\State {$u_t \leftarrow \Call{Policy}{b_t}$}
\State{$\xi_{t+1} \leftarrow \Call{Dynamics}{\xi_t,u_t}$}
\State{$o_{t} \leftarrow \Call{ReceiveMeasurements}{\xi_{t+1}}$}
\State{$b_{t+1} \leftarrow \Call{UKF}{b_t,u_t,o_{t}}$}
\EndFor
\end{algorithmic}
\end{algorithm}
\vspace{-3mm}

\subsection{Probabilistic Bounding Heuristic}

MPC is popular in part because of its ability to incorporate feasibility constraints. In particular, for many problems it can guarantee \emph{persistent feasibility}, that is, that the state will stay in a specified region in all future steps \cite{bertsekas1995dynamic}. For a system with time-varying and normally distributed parameters, the possible disturbances to the system are unbounded. Thus, no controller can guarantee feasibility. To approximate feasibility, a heuristic was developed to select appropriate actions at each state node to keep the norm of the next state within a desired bound. The heuristic consists of three steps: computing confidence regions, truncating samples, and checking the norm of the next state by propagating uniformly sampled actions. If this heuristic is followed, the norm of the next state is guaranteed to lie within the desired confidence region $\beta_\text{des}$ with probability equal to the confidence level. Similar heuristics that use confidence intervals and regions when estimating partially identified parameters have been studied in the past \cite{stoye2009more,cui2003empirical}. 

The norm of the next state can be written as
%
\begin{equation}
\Vert x_{k+1} \Vert = \Vert f(x_k, \theta_k, u_k) + w_k \Vert
\text.
\end{equation}
An upper bound for this next state norm is computed by finding the probabilistic worst case given the process noise distribution and hyperstate distribution estimated by the UKF. Since the process noise is additive, it can be separated to create a conservative norm approximation
\begin{equation}
\Vert x_{k+1} \Vert \leq \beta (b_k, u_k) \text{,}
\end{equation}
defined as
\begin{align}
\beta (b, u) &= \beta_b(b, u) + \beta_w \\
\beta_b(b, u) &= \max_{\xi \in \conreg(\hat{\xi}, \Sigma^\xi, \alpha)} \Vert f(\xi,u) \Vert \\
\beta_w &= \max_{w \in \conreg(0, \Sigma^w, \alpha)} \Vert w \Vert \text{,}
\end{align}
where $\conreg(\mu, \Sigma, \alpha)$ is the confidence region ellipsoid defined in (\ref{eq:confreg}).

Confidence regions are computed from the given distributions to find the components of the upper bound corresponding to the belief over the hyperstate $\beta_{b}$ and process noise $\beta_w$. This assumes that the UKF hyperstate belief and process noise are true distributions. Samples within the confidence region are found by uniformly sampling $n_b$ points inside a spheroid of the appropriate dimension. These samples are transformed to uniform samples inside a confidence region $S_c$ following the computation of the approximate confidence region ellipsoid in \Cref{sec:approxconfreg}. 

\begin{algorithm}
\begin{algorithmic}[1]
\caption{Probabilistic bounding heuristic}
\label{probalg}
\Require $b_0, w, \beta_{des}, \alpha, n_u, n_b$
\State{$\Sigma^w \leftarrow \Call{ConfidenceRegion}{w, \alpha}$}
\State{$\Lambda^w \leftarrow \Call{Eigendecomposition}{\Sigma^w}$}
\State{$\beta_{w} \leftarrow \underset{\lambda \in \Lambda^w}{\max} \sqrt{\vert \lambda \vert} $}
\State{$\Sigma^c \leftarrow \Call{ConfidenceRegion}{b_0, \alpha}$}
\State{$S_c \leftarrow \Call{EllipsoidSample}{\Sigma^c,n_b}$}
\State{$S_b \leftarrow \Call{TruncateParameters}{S_c}$}
\For{$i \in [1, \dots, n_u]$}
\State{$u_i \leftarrow \Call{UniformSample}{U}$}
\State{$S_{b'} \leftarrow \Call{PropagateDynamics}{S_b,u_i}$}
\State{$\beta_{b} \leftarrow \underset{s \in S_{b'}}{\max} \Vert s \Vert$}
\If{$\beta_w + \beta_{b} \leq \beta_{des}$} 
Break
\EndIf
\EndFor
\State\Return $u_i$ 
\end{algorithmic}
\end{algorithm}
The samples within the confidence region are computed for both the process noise and belief over the hyperstate.
$\beta_w$ is simply the square root of the eigenvalue in the confidence region with the largest absolute value. The confidence region samples for the hyperstate belief are checked to ensure they meet the lower bounds on unknown parameter values defined in the model before being included in the sample set.
The truncated samples are propagated through the dynamics with a randomly selected action to compute $\beta_{b}$. 

The bound estimates $\beta_{b}$ and $\beta_w$ are summed and compared to the desired bound.
If the uniformly sampled action will keep the system within the desired bound, the action is added to the search tree. Otherwise, a new action is drawn and the hyperstate belief samples are propagated until the desired bound is met, or a specified number of iterations $n_u$ have occurred. Thus, as long as an action likely exists to keep the system within the desired bounds, the system will stay within that bound with probability at least equal to the given confidence level. This process is summarized in Algorithm \ref{probalg}.

\section{Simulation and Results} \label{sec:results}

A model of a robot performing planar manipulation was used to test the simultaneous estimation and control capability of MCTS, QMDP-TS, and MPC.

\subsection{Planar Manipulation Model}

\begin{figure}
\centering
\includegraphics[width=0.3\textwidth]{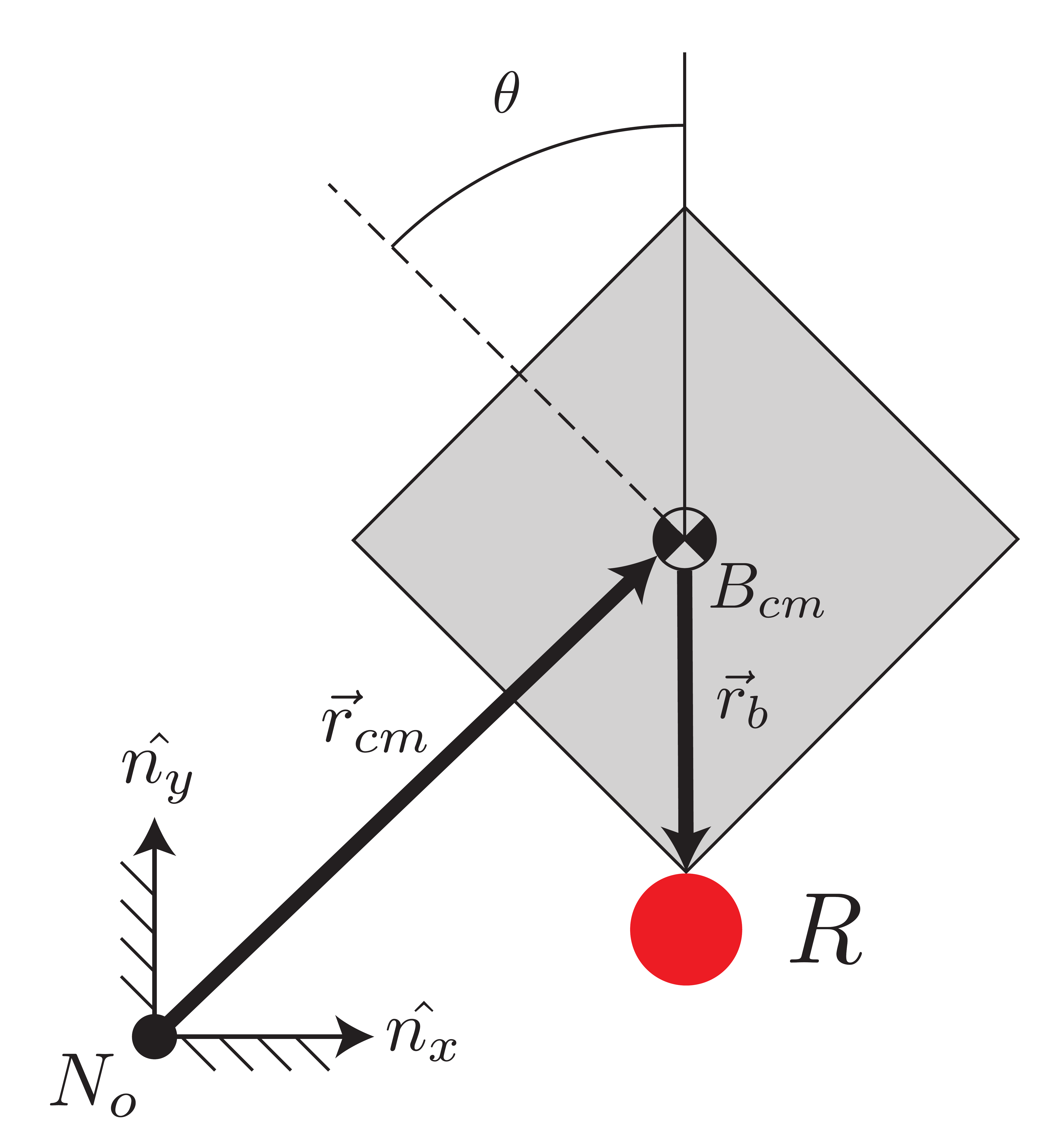}
\caption{Schematic of planar  manipulation task}
\label{fig:2Dmanip}
\end{figure}

We consider an agent $R$ pushing a box $B$ in the plane, where the agent may apply an arbitrary force $F$ in the $x$ and $y$ directions, in addition to a torque ${T}$. This problem, along with relevant parameters and variables used to describe the system state, is illustrated in Figure \ref{fig:2Dmanip}.
The state-space form of the system uses the state vector $x_k = \left[ p_{x,k}, p_{y,k}, p_{\theta,k}, v_{x,k}, v_{y,k}, v_{w,k} \right]^T$, corresponding to the linear and angular positions and velocities of $B$ in the global frame $N$. The linear and angular accelerations are given as $a_k = \left[ a_{x,k},  a_{y,k},  a_{\alpha,k} \right]^T$. The system has time-varying, unknown parameters $\theta_k = \left[ m_k, \mu_{v,k}, J_k, r_{b,x,k}, r_{b,y,k} \right]^T$ which represent the mass, linear friction, inertia, and distance from the center of mass of the box to the robot with respect to the $x$ and $y$ directions. The unknown parameters are limited to enforce a lower bound that prevents non-physical values such as negative mass. This bound truncates the Gaussian distribution over the unknown parameters.

We can describe the dynamics of the system about its center of mass, $B_{cm}$. These are given as
\begin{equation}
{F}_{B,x} = {F}_x - \mu_v {v}_x =  m {a}_x
\label{force_eq_x}
\end{equation}
\begin{equation}
{F}_{B,y} = {F}_y - \mu_v {v}_y =  m {a}_y
\label{force_eq_y}
\end{equation}
\begin{equation}
{T}_B= {T} + \vec{r}_b \times \vec{F}_{B} = J {a}_{\alpha}
\text.
\label{torque_eq}
\end{equation}
The system in discrete time with step size of $\Delta t$ is
\begin{eqnarray}
	\vec{v}_{k+1} = \vec{a}_k  \Delta t+ \vec{v}_k\\
	\vec{p}_{k+1} = \vec{v}_k \Delta t + \vec{p}_k\text.
\end{eqnarray}
Rewriting (\ref{force_eq_x}), (\ref{force_eq_y}), and (\ref{torque_eq}) in state-space form gives
\begin{equation}
x_{k+1} = f(x_k,u_k) = A x_k \\ + B(\theta_k, p_{\theta,k}) u_k + w_k\text,
\end{equation}
where
\begin{equation}
A = \left[ \begin{array}{c|c}
\mathbb{I}^{3x3} & \mathbb{I}^{3x3} \Delta t  \\
\hline \\ [-0.8em]
\huge 0^{3x3}  & \mathbb{I}^{3x3}
\end{array} \right] \end{equation}
\begin{equation}
B=  \left[\begin{array}{ccc}
\multicolumn{3}{c}{\Huge 0^{3x3}} \\
\hline \\[-0.8em] \frac{\Delta t}{m_k} & 0 & 0\\
0 & \frac{\Delta t}{m_k} & 0\\
B^{3,1} & B^{3,2} & \frac{\Delta t}{J_k} \end{array}\right]
\end{equation}
\begin{equation}
B^{3,1} =  \frac{\Delta t}{J_k}(\cos(p_{\theta,k})r_{b,y,k} + \sin(p_{\theta,k})r_{b,x,k})
\end{equation}
\begin{equation}
B^{3,2} =  \frac{\Delta t}{J_k}(\cos(p_{\theta,k})r_{b,x,k} - \sin(p_{\theta,k})r_{b,y,k})
\end{equation}
\begin{equation}
u_k = \left[ \begin{array}{c}
F_{x,k} \\ F_{y,k} \\ T_k \end{array}\right]\text,
\end{equation}
with $m$ as the mass of $B$ and $J$ is $B$'s moment of inertia.

For a robot with noisy sensors which measure its position, velocity, and acceleration in $N$, the observation model is
\begin{equation}
	y_k = h(x_k,u_k) + v_k
\end{equation}
where $y_k = \left[ p_{x,k},  p_{y,k},  p_{\theta,k},  v_{x,k},  v_{y,k}, v_{w,k}, a_{x,k},  a_{y,k},  a_{\alpha,k} \right]^T$.

The measurement functions are given by
\begin{align}
\vec{p} &= \vec{r}_{cm} + \vec{r}_b \\
\vec{v} &= \vec{v}_{cm} + {v}_{w,k} \times \vec{r}_b\\
%
{a}_{\alpha,k} &= \frac{{T}_B}{J_k} \\
\left[{a}_{x,k},  {a}_{y,k} \right] &= \frac{\vec{F}_B}{m_k}+ {a}_{\alpha,k} \times \vec{r}_b + {v}_{w,k}\times({v}_{w,k} \times \vec{r}_b)\text.
\end{align}

The goal for this scenario is to reach a stationary goal at the origin with an orientation in the positive $x$-direction. It is given a random initial position, orientation, velocity, and parameters drawn from the initial parameter distribution.

\subsection{Implementation Details}

For the evaluation tests, a simulation consists of 50 steps, a step duration of 0.1 seconds, and a total of 100 trials for each simulation condition. The initial parameter distribution in all simulations is a normal distribution with a mean of 1 and variance of 0.5. The magnitude of control inputs is limited to a value of 5.0 which allows a simulation of MPC with no process noise to reach the goal state in less than half of the 50 steps in each simulation, providing enough control effort to possibly reach the goal state under uncertain conditions. Measurement noise was not included to isolate the effects of process noise and the lower bound on unknown parameters. 

The reward function is linear with a weighted $L_1$ norm penalty for the position, speed, and control effort
\begin{equation}
R_{L_1}(x_k,u_k) = \left[\begin{array}{cc} R_{pos} & 0 \\ 0 & R_{vel} \end{array}\right] \vert  x_k \vert + R_u \vert u_k \vert \text.
\end{equation}
The values for $R_{pos}$, $R_{vel}$, and $R_u$ are $-2.5$, $-50$, and $-0.3$, respectively. 

The MCTS and QMDP-TS implementations use a discount factor of 0.99. In the tree search, the next actions to be explored from a state node are selected by an epsilon greedy strategy. A random action is taken with probability 0.8, otherwise an MPC policy is computed using a state randomly sampled from the belief. This biases the actions to policies that generally perform well when accurate parameter estimates are available. The total number of nodes  in the tree search is limited to 3000 for all simulations other than the bounded heuristic simulations which uses 300 total nodes. The bounding heuristic limited the maximum number of actions, $n_u$ to 50.
The MCTS with DPW implementation is from the POMDPs.jl package \cite{egorov2017pomdps}. The optimization in the MPC controller is solved with the Convex.jl package \cite{udell2014convex}.

\subsection{Cross-Entropy for Tuning Solver Parameters}

The cross-entropy method is used offline to optimize the hyperparameters for MCTS and QMDP-TS.
The hyperparameters are the number of actions sampled at each state node, the number of states sampled at each action node, the depth of the tree, and the exploration constant.
The parameters for the DPW search were constrained by fixing $\delta = \frac{1}{30}$ so that only $k$ is free to be optimized to control the number of child states in (\ref{eq:dpw}).

The initial parameter distribution is chosen heuristically with a mean of $\mu_{ce} = \left[20, 20, 10, 20 \right]$ and covariance of $\Sigma_{ce} = \diag(64, 64, 16, 81)$ for the respective hyperparameters.
The samples from this distribution are rounded to the closest integer value with a minimum value enforced at 1.
For the cross-entropy optimization, the population size is 50, 10 elite samples are used to fit the next distribution, the iteration limit is 25, and the optimization is ended when the maximum eigenvalue drops below 3.

The parameters for all test cases are the cross-entropy results for the condition with process noise variance 0.01. The resulting mean values for the MCTS hyperparameters were $\mu = \left[22, 5, 12, 27 \right]$.
The planning horizon of MPC was selected to match the optimized MCTS depth of 12.

\subsection{Performance}

\begin{figure}
\begin{subfigure}{\linewidth}
  \centering
  \begin{tikzpicture}[]
\begin{axis}[legend pos = {south east}, ylabel = {$y$ Position (m)}, xmin= {-3}, xmax = {4}, ymin = {-6}, ymax = {2}, xlabel = {$x$ Position (m)}, every axis plot/.append style={line width=1pt}]\addplot+ [sharp plot]coordinates {
(1.0, 1.0)
(0.6703168126597772, 0.8959088367212549)
(0.8913297037288662, 0.20262726555010135)
(0.4055420176646176, -0.15275731235105638)
(0.5128502578185516, -0.005743911060092408)
(0.5236399413003602, 0.35778216387369205)
(0.900999207074452, 0.6938776433161671)
(0.9439808069213232, 0.6805417888911922)
(1.2179204318620025, 0.9397275894165776)
(0.6730796050540835, 0.5541027222553438)
(0.03058418855152397, 0.012343226710423288)
(-0.557028226176739, -0.4845847134915796)
(-1.0438181249419896, -0.8641387415924908)
(-1.645677507447592, -1.11212784707843)
(-2.008399431762215, -1.6267403436155303)
(-2.396026435904339, -1.8129923152176155)
(-2.7755867993161525, -1.9530170390888473)
(-2.936184175566359, -2.33860773584502)
(-2.233392559244177, -2.0258947637938594)
(-2.537888577765801, -2.280177592760045)
(0.2285057424667568, -1.60527408572004)
(1.105363657708976, 0.4821841562597688)
(1.2274770086184132, -1.5381221283158049)
(0.004779956023032543, -2.0145754009190795)
(0.049045999220156726, -1.41111776553184)
(-0.3382023145883179, -1.4326111424862595)
(-0.5637556900970633, -1.277919606733272)
(-0.7594338757422248, -1.4594213265168263)
(0.21997703500731292, -1.0265529886338804)
(1.240138953800502, 0.07291170667825145)
(1.773696770929084, 0.31278484327103506)
(0.987481911219874, -0.23776713729739757)
(0.6818101186132782, -0.12547772497143583)
(0.4485110622072001, -0.34102558303031993)
(-0.18828497596454188, -0.865936637870227)
(-0.8466121153405161, -1.3082235691411754)
(-0.4250550189580948, -0.23895887943309513)
(-0.6119584428985043, 0.23911075616107785)
(0.208754043415992, 0.4201210264520423)
(-0.7768666704794417, -0.3282783331984101)
(-0.8293440556665719, -0.5837087966226044)
(-0.028828433755725625, 0.12099819402149312)
(0.07237603133028729, 1.064852351013174)
(-0.2380643581685149, 0.6331292023549722)
(0.6208306134653162, -0.4400965168205362)
(0.37467290842495954, -0.7355961699237888)
(0.2765478427728447, -0.5335451817230629)
(-0.13649887331926208, -0.903959585271239)
(-0.8054408624774788, -0.5698134470894745)
(-1.455790726205015, -0.5357044263453685)
};
\addlegendentry{True State}
\addplot+ [sharp plot]coordinates {
(1.4562329122151991, 1.0708872395980835)
(0.46594408605752946, 0.2430351386994951)
(1.454407554713484, -0.12202380357474828)
(0.7017101973588565, -0.16367206823792255)
(0.8231627950764945, -0.06517523541340473)
(1.0463373213533946, 0.28372904580243974)
(0.983213967966245, 0.8044942915786275)
(0.8769215795636915, 0.6214421091221728)
(1.1382931362919473, 0.9235166116201357)
(0.5690670834297376, 0.3254525080943145)
(0.03887059755175095, -0.06323048543458312)
(-0.6243000463782439, -0.6215534718388586)
(-0.9983223329761886, -0.95000781921126)
(-1.3268056738693648, -1.163329094978879)
(-1.9017760451258827, -1.5246703744270704)
(-2.218903259037205, -2.0420889598821472)
(-2.397777463982, -1.9524948504772024)
(-2.930956855390398, -2.268144780457674)
(-2.447931568576222, -1.929891788038661)
(-2.4811242887910945, -2.388201348738093)
(0.3474870856781649, -1.506532825468781)
(1.1119872157539863, 0.34976031512277905)
(1.2932273723975007, -1.4929972461373804)
(-0.019138251001838524, -1.852504539353601)
(-0.06081759830288933, -1.525881229085284)
(-0.3231842975332735, -1.4620960232522044)
(-0.3686626899977117, -1.2849720541678218)
(-0.7888013161530555, -1.380063825345747)
(0.19727767043423605, -1.2154949675751396)
(1.030081893084914, -0.03590007663169027)
(1.7721759513583626, 0.3653435874529323)
(0.890039792867699, -0.32306977560056793)
(0.864963483781065, -0.11377485461310212)
(0.5408902846317232, -0.19519551307657887)
(0.08826366131805606, -0.6509922797056455)
(-0.6636895160279317, -1.2841527726631978)
(-0.41169977395192625, -0.35685370995913956)
(-0.6298307491227944, 0.24660221243221864)
(0.29376318376361665, 0.5370359722096294)
(-0.6533381574804648, -0.26837214022190353)
(-0.9252946494614236, -0.4983487675046767)
(0.014926664765158124, 0.18305951217003183)
(-0.009241070544894573, 1.2453799550632754)
(-0.2865300454103884, 0.5684538958160439)
(0.8653404772887328, -0.6014867615787258)
(0.5794383559581695, -0.7558689582729181)
(0.2465258300358684, -0.5343124216900952)
(-0.03920701597344205, -0.6028542675615363)
(-0.9104564610904738, -0.6793730349315658)
(-1.3091134597615106, -0.5857450277546524)
};
\addlegendentry{State Estimate}
\end{axis}

\end{tikzpicture}
  \subcaption{MCTS example trajectory}
  \label{fig:mcts_traj}
\end{subfigure}%
\par\bigskip 

\begin{subfigure}{\linewidth}
  \centering
  \begin{tikzpicture}[]
\begin{axis}[legend pos = {south east}, ylabel = {$y$ Position (m)}, xmin= {-3}, xmax = {4}, ymin = {-6}, ymax = {2}, xlabel = {$x$ Position (m)}, every axis plot/.append style={line width=1pt}]\addplot+ [sharp plot]coordinates {
(1.0, 1.0)
(0.4466025903149526, 0.3920030904228507)
(-0.2028215209919288, -0.03349927266461905)
(-0.7318050248580241, 0.701903715887093)
(-0.6780071118846381, 0.8079166267332222)
(-1.065931672796511, 0.2855135369705155)
(-1.420621338559277, -0.1592815705375804)
(-1.2032309176186815, -0.05420277426894991)
(-1.0609737037995814, -0.13222126522167)
(-1.0180208417309404, -0.013823602920764788)
(-1.2737597799842098, -0.34994193457268474)
(-1.1584105102424644, -0.85272741371079)
(-0.7758971733378361, -1.1617414147263885)
(-1.0801902556777965, -1.5319656832701056)
(-1.4758137887189071, -1.7712241884544018)
(-1.8425088868951291, -2.009795431754937)
(-2.273679116449891, -2.2437626325056095)
(-2.583747742551903, -2.3017843343800983)
(-1.9599779802358732, -1.6594544865983756)
(-1.2465424756424963, -1.9748087088107256)
(-0.5883777000793908, -2.2257458798722927)
(-1.0213294965785802, -1.6422837003404056)
(-0.6778690900781457, -2.117375615712388)
(-0.3111752812694117, -2.4027283973984637)
(0.06637362863086625, -2.4707322345049603)
(0.5438082415795877, -2.7220877284780753)
(0.843765682695908, -2.19711314365673)
(1.2790366927150547, -1.6699809870619966)
(0.6916660247561088, -1.2290689317556638)
(0.4216572853975231, -0.8305835584330818)
(-0.0737805887778848, -0.5301411856097836)
(-0.5438170592084209, -0.11824718080084284)
(-0.6789128196192081, -0.0997369455021725)
(-0.5843388469076494, -0.18915608893236796)
(-0.5002051588309955, -0.13955783024277216)
(-0.5157386665102225, -0.16082221556695797)
(-0.4757714057279181, -0.09805981796855724)
(-0.5513038608587951, -0.19698329712189735)
(-0.5240795587095155, -0.06505541605131004)
(-0.41159532954181954, -0.1843308373821296)
(-0.5206819186966207, -0.06774060560964884)
(-0.4870798548697819, -0.21616940158844303)
(-0.3721041163080757, -0.36553757799328257)
(-0.1482811756660398, -0.4562584616764519)
(-0.15488264997264317, -0.5555417769875455)
(-0.14818468753310157, -0.419285553740553)
(-0.009057330825645, -0.420437937631424)
(-0.07405495591623415, -0.334497101472045)
(-0.1308617788600121, -0.3221041414619757)
(-0.05730705756824818, -0.2986023136044581)
};
\addlegendentry{True State}
\addplot+ [sharp plot]coordinates {
(0.5533003008716484, 1.4364733661896525)
(0.6133952783473231, 0.8817555773744349)
(0.8545424638207496, -0.14434962967537213)
(-0.5833511259697133, -0.5581144410303244)
(-0.6611303193653256, -0.03900136902664947)
(-1.4717568678306137, 0.8373742702575444)
(-1.2527732485693344, 0.382478012194516)
(-0.7629808061046194, -0.037124244423471675)
(-0.5986891615521448, 0.12416956336703353)
(-0.8395415595562237, -0.001993159358069277)
(-1.3976057246641005, -0.492687605278014)
(-1.2510811383484337, -0.7287581139704384)
(-0.8915425281670516, -1.2516431963253454)
(-1.740540718193376, -1.4072670470024295)
(-1.5789966652271477, -1.7577980506675697)
(-0.975132014571664, -1.2919732907612873)
(-2.248692356780242, -2.0105548417467753)
(-2.7583429314970855, -2.206064569275044)
(-1.3542563563773795, -0.5380265330747579)
(-1.010197008882066, -1.9816922378180215)
(-1.0247394569543786, -2.5589021159397545)
(-2.31786286094532, -2.0510892135402297)
(-1.2469037798806475, -2.079982743705)
(0.028032094284458617, -3.926548789590457)
(-0.11991138917287424, -4.371123073251998)
(2.431326845813435, -3.4736063873701744)
(2.6643974306083575, -2.3233149330339415)
(3.9874467364067456, 0.9584105426173152)
(3.1420482345232617, 1.445870193648291)
(0.8928196984662722, 0.15348773686462902)
(0.5501307184807285, -5.418012176948945)
(10.090265561250822, -14.1343787745367)
(11.444428232678518, -22.54862714641081)
(11.444428232678518, -22.54862714641081)
(11.444428232678518, -22.54862714641081)
(11.444428232678518, -22.54862714641081)
(11.444428232678518, -22.54862714641081)
(11.444428232678518, -22.54862714641081)
(11.444428232678518, -22.54862714641081)
(11.444428232678518, -22.54862714641081)
(11.444428232678518, -22.54862714641081)
(11.444428232678518, -22.54862714641081)
(11.444428232678518, -22.54862714641081)
(11.444428232678518, -22.54862714641081)
(11.444428232678518, -22.54862714641081)
(11.444428232678518, -22.54862714641081)
(11.444428232678518, -22.54862714641081)
(11.444428232678518, -22.54862714641081)
(11.444428232678518, -22.54862714641081)
(11.444428232678518, -22.54862714641081)
};
\addlegendentry{State Estimate}
\end{axis}

\end{tikzpicture}
  \subcaption{Cautious MPC example trajectory}
  \label{fig:mpc_traj}
\end{subfigure}
\par\bigskip 

\caption{Example trajectories of the true state and state estimates from (a) MCTS and (b) cautious MPC for a process noise variance of 0.01 and unknown parameter lower bound of 0.05. The linear interpolation between points do not reflect the dynamics of the system.}
\label{fig:traj}
\end{figure}
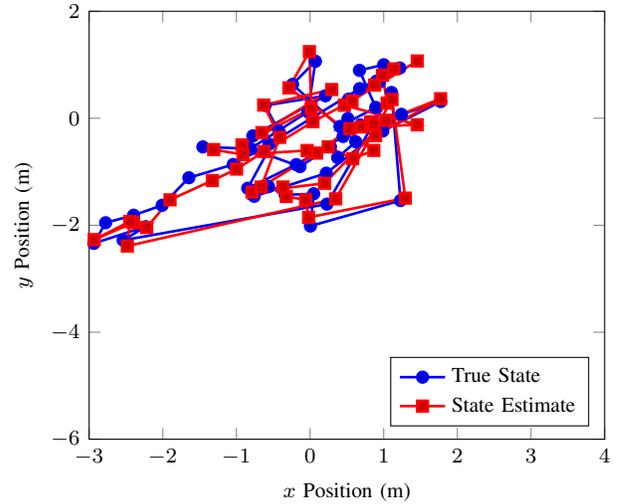
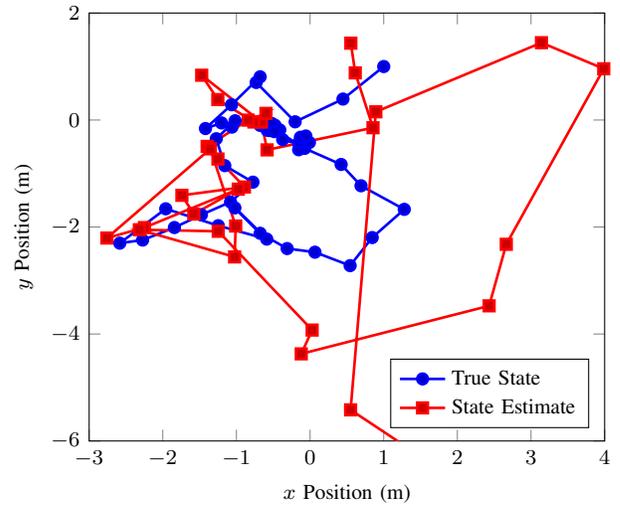

First, single trajectories from the MCTS and MPC solution methods are shown to illustrate the qualitative difference between their behavior before statistical results are presented.
A single simulation trajectory from the MCTS simulations is visualized in Fig. \ref{fig:mcts_traj}, and a trajectory from the MPC simulations is shown in Fig. \ref{fig:mpc_traj}.
The MCTS true and estimated state trajectory match well. The MPC state estimate diverges from the true state as the number of steps increases. MCTS keeps the state closer to the origin by gathering information to achieve a better estimate.

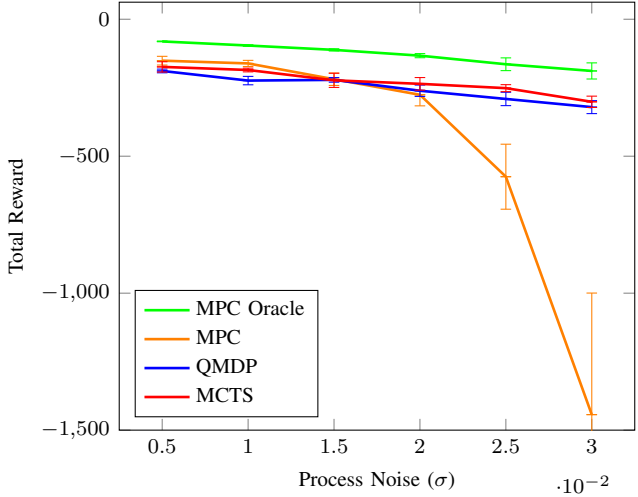
\begin{figure}
\centering
\begin{tikzpicture}[]
\begin{axis}[legend pos = {south west}, ylabel = {Total Reward}, ymin = {-1500}, xlabel = {Process Noise ($\sigma$)}, every axis plot/.append style={line width=1pt}]\addplot+ [
mark = {none}, green, error bars/.cd,
x dir=both, x explicit, y dir=both, y explicit]
table [
x error plus=ex+, x error minus=ex-, y error plus=ey+, y error minus=ey-
] {
x y ex+ ex- ey+ ey-
0.005 -81.05874658790088 0.0 0.0 1.2374140225187984 1.2374140225187984
0.01 -95.97688756376672 0.0 0.0 2.8211780994052997 2.8211780994052997
0.015 -111.86403051675636 0.0 0.0 4.073206656986921 4.073206656986921
0.02 -132.96036969267877 0.0 0.0 7.548210831120093 7.548210831120093
0.025 -164.33443168164436 0.0 0.0 23.281817273809803 23.281817273809803
0.03 -188.7479191280565 0.0 0.0 29.44360768019351 29.44360768019351
};
\addlegendentry{MPC Oracle}
\addplot+ [
mark = {none}, orange, error bars/.cd, 
x dir=both, x explicit, y dir=both, y explicit]
table [
x error plus=ex+, x error minus=ex-, y error plus=ey+, y error minus=ey-
] {
x y ex+ ex- ey+ ey-
0.005 -151.26850182188969 0.0 0.0 15.894862430153156 15.894862430153156
0.01 -161.34158266795558 0.0 0.0 11.223535518429925 11.223535518429925
0.015 -219.0284477666659 0.0 0.0 22.62095436182215 22.62095436182215
0.02 -275.84641265841725 0.0 0.0 40.32982152492319 40.32982152492319
0.025 -574.8314714604 0.0 0.0 118.6499415419 118.6499415419
0.03 -1443.4098372366 0.0 0.0 443.8238423882 443.8238423882
};
\addlegendentry{MPC}
\addplot+ [
mark = {none}, blue, error bars/.cd, 
x dir=both, x explicit, y dir=both, y explicit]
table [
x error plus=ex+, x error minus=ex-, y error plus=ey+, y error minus=ey-
] {
x y ex+ ex- ey+ ey-
0.005 -188.7895432036004 0.0 0.0 6.780693176090258 6.780693176090258
0.01 -223.96657888790415 0.0 0.0 15.34673977956612 15.34673977956612
0.015 -221.3875692216383 0.0 0.0 8.446693201792892 8.446693201792892
0.02 -261.0156835647079 0.0 0.0 20.699497916785035 20.699497916785035
0.025 -290.9851354631 0.0 0.0 24.4132256521 24.4132256521
0.03 -320.803251838134 0.0 0.0 23.72210464558928 23.72210464558928
};
\addlegendentry{QMDP}
\addplot+ [
mark = {none}, red, error bars/.cd, 
x dir=both, x explicit, y dir=both, y explicit]
table [
x error plus=ex+, x error minus=ex-, y error plus=ey+, y error minus=ey-
] {
x y ex+ ex- ey+ ey-
0.005 -173.9441441284 0.0 0.0 18.8549000238 18.8549000238
0.01 -185.3025090199812 0.0 0.0 6.8501434236061485 6.8501434236061485
0.015 -223.0096929444 0.0 0.0 25.876384563 25.876384563
0.02 -235.8600484738 0.0 0.0 22.9391979613 22.9391979613
0.025 -251.58232727082455 0.0 0.0 12.75793378375132 12.75793378375132
0.03 -301.4418364422 0.0 0.0 20.6926626767 20.6926626767
};
\addlegendentry{MCTS}
\end{axis}

\end{tikzpicture}
\caption{Total rewards averaged over 100 simulations for compared methods at various levels of process noise.}
\label{fig:noise_plot}
\end{figure}

Simulation results for a range of process noise levels and a fixed parameter lower bound of 0.0625 are shown in Fig. \ref{fig:noise_plot}. This lower bound prevents any of the unknown parameter values from becoming smaller than 0.0625 due to the added process noise. The error bars represent the standard error of the mean. For all methods, total rewards generally decrease as process noise increases. The MPC oracle is an upper bound computed by allowing the agent to fully observe the time-varying parameter values. All methods other than the oracle achieve similar rewards for conditions with little process noise where the uncertainty is small enough that the model estimate does not require exploration to be accurate. MCTS and QMDP-TS outperform MPC by a large margin for higher process noise levels. This indicates that accounting for uncertainty is beneficial in this problem. While MCTS achieves higher rewards than QMDP in higher process noise levels, the difference is often within the standard error of the mean. Thus, accounting for uncertainty beyond one step offers little improvement for an unknown parameter lower bound of 0.0625.

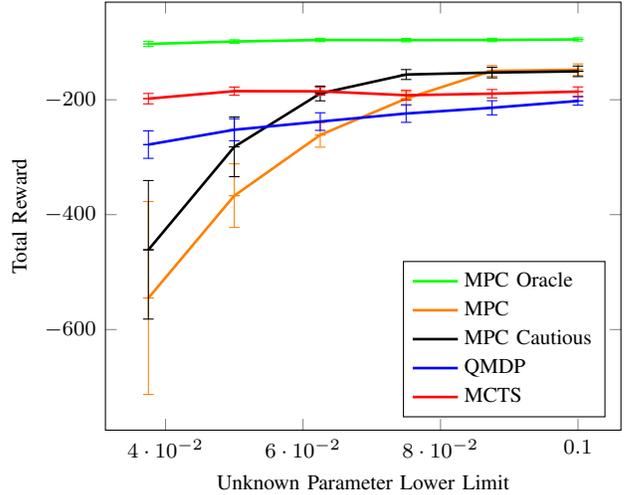
\begin{figure}
\centering
\begin{tikzpicture}[]
\begin{axis}[legend pos = {south east}, ylabel = {Total Reward}, xlabel = {Unknown Parameter Lower Limit},every axis plot/.append style={line width=1pt}]\addplot+ [
mark = {none}, green, error bars/.cd, 
x dir=both, x explicit, y dir=both, y explicit]
table [
x error plus=ex+, x error minus=ex-, y error plus=ey+, y error minus=ey-
] {
x y ex+ ex- ey+ ey-
0.0375 -102.8501124803 0.0 0.0 4.578620199 4.578620199
0.05 -98.66790903125263 0.0 0.0 3.022568954690243 3.022568954690243
0.0625 -95.97688756376672 0.0 0.0 2.8211780994052997 2.8211780994052997
0.075 -96.31347370640584 0.0 0.0 2.98561678842572 2.98561678842572
0.0875 -95.99604654921905 0.0 0.0 2.947928855348396 2.947928855348396
0.1 -95.04100055548446 0.0 0.0 2.7754689513206148 2.7754689513206148
};
\addlegendentry{MPC Oracle}
\addplot+ [
mark = {none}, orange, error bars/.cd, 
x dir=both, x explicit, y dir=both, y explicit]
table [
x error plus=ex+, x error minus=ex-, y error plus=ey+, y error minus=ey-
] {
x y ex+ ex- ey+ ey-
0.0375 -545.0802244318 0.0 0.0 167.9305020035 167.9305020035
0.05 -366.9564310704 0.0 0.0 55.3850991045 55.3850991045
0.0625 -261.341582668 0.0 0.0 21.2235355184 21.2235355184
0.075 -197.9704183165 0.0 0.0 11.8550364433 11.8550364433
0.0875 -149.72526924137196 0.0 0.0 9.33744082867713 9.33744082867713
0.1 -147.72641521575028 0.0 0.0 10.151490165644152 10.151490165644152
};
\addlegendentry{MPC}
\addplot+ [
mark = {none}, black, error bars/.cd, 
x dir=both, x explicit, y dir=both, y explicit]
table [
x error plus=ex+, x error minus=ex-, y error plus=ey+, y error minus=ey-
] {
x y ex+ ex- ey+ ey-
0.0375 -461.1846289688 0.0 0.0 120.5902015509 120.5902015509
0.05 -281.9816822841 0.0 0.0 51.9579633018 51.9579633018
0.0625 -189.0356826796 0.0 0.0 12.8251744045 12.8251744045
0.075 -156.09903205799606 0.0 0.0 8.65714952916948 8.65714952916948
0.0875 -152.64420839931253 0.0 0.0 9.182077727614503 9.182077727614503
0.1 -150.51331252773743 0.0 0.0 9.172761173564503 9.172761173564503
};
\addlegendentry{MPC Cautious}
\addplot+ [
mark = {none}, blue, error bars/.cd, 
x dir=both, x explicit, y dir=both, y explicit]
table [
x error plus=ex+, x error minus=ex-, y error plus=ey+, y error minus=ey-
] {
x y ex+ ex- ey+ ey-
0.0375 -278.0755923304 0.0 0.0 23.8882232142 23.8882232142
0.05 -252.2284760271 0.0 0.0 19.1583073895 19.1583073895
0.0625 -237.9665788879 0.0 0.0 15.3467397796 15.3467397796
0.075 -223.96657888790415 0.0 0.0 15.34673977956612 15.34673977956612
0.0875 -213.96879234910415 0.0 0.0 12.3346475626124 12.3346475626124
0.1 -202.0923448212349 0.0 0.0 7.347233051946819 7.347233051946819
};
\addlegendentry{QMDP}
\addplot+ [
mark = {none}, red, error bars/.cd, 
x dir=both, x explicit, y dir=both, y explicit]
table [
x error plus=ex+, x error minus=ex-, y error plus=ey+, y error minus=ey-
] {
x y ex+ ex- ey+ ey-
0.0375 -198.0895604242 0.0 0.0 9.0735944641 9.0735944641
0.05 -185.04107950542553 0.0 0.0 7.204036851306246 7.204036851306246
0.0625 -185.3025090199812 0.0 0.0 6.8501434236061485 6.8501434236061485
0.075 -192.09938415906032 0.0 0.0 8.516557045951995 8.516557045951995
0.0875 -189.47928005560044 0.0 0.0 7.28457192362349 7.28457192362349
0.1 -185.78535206732872 0.0 0.0 8.197576572272434 8.197576572272434
};
\addlegendentry{MCTS}
\end{axis}

\end{tikzpicture}
\caption{Total rewards averaged over 100 simulations for the compared methods with various lower bounds on unknown parameters for a process noise variance of 0.01.}
\label{fig:lim_plot}
\end{figure}

The lower bound of the unknown parameters also has a significant effect on performance.
A smaller lower bound allows smaller mass, friction coefficient, and other parameter values. These smaller parameter values may result in a larger change in the state when the same control inputs are applied. As the lower bounds decrease, the total reward of MPC decreases the most, with QMDP-TS rewards decreasing to a lesser degree in Fig. \ref{fig:lim_plot}. 
The reward for MCTS remains approximately the same across all conditions. Accounting for additional uncertainty useful information that maintained performance when the unknown parameters had lower possible bounds. 
The MPC oracle reward remains relatively constant across conditions, indicating that the difficulty of the reaching the goal with full knowledge does not change significantly when the bound changes, and the difference in performance is due to the effects of uncertainty.
An additional MPC simulation, cautious MPC, uses an artificially inflated process noise in the UKF to try to better maintain the consistency of the filter in the presence of significant nonlinearity. Cautious MPC improves the total reward in comparison to standard MPC, but matches the decline in performance for lower parameter bounds.

\begin{figure}
\centering
\input{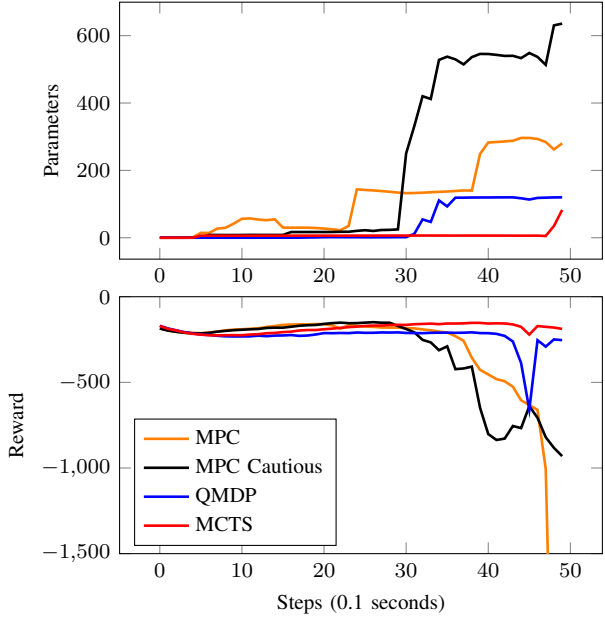}
\caption{Mean absolute error of the unknown parameters and the rewards at each simulation step averaged over 100 simulations for a process noise variance of 0.01 and unknown parameter lower bound of 0.05.}

\label{fig:param_varying_plot}
\end{figure}

Time-domain plots give intuition about the behavior of the methods.
The error in the unknown parameters and rewards are averaged over all simulations and plotted for each time step in Fig. \ref{fig:param_varying_plot}.
MPC shows inconsistent growth in the unknown parameter error beginning within the first 5 steps of simulation. Cautious MPC has little unknown parameter error until the second half of the simulation when the error increases quickly. This delayed but large error is likely due to the increased process noise, preventing cautious MPC from converging to an accurate estimate. The parameter error for QMDP-TS increases to a smaller error than either MPC method. MCTS achieves the smallest parameter error in all cases with only a significant increase in the last few steps of the simulation. The growth of the parameter error is reflected in worsening rewards accrued over the time steps.

The simulations were analyzed to determine if the performance of any methods are affected by UKF divergence. The divergence criteria is met if any component in the hyperstate error is larger than 5 times the square root of the corresponding eigenvalue of the covariance. The hyperstate error was rotated by the eigenvectors of the covariance vector before the comparison. None of the methods met this criteria, indicating the UKF estimation was not a major factor in the performance.

\subsection{Probabilistic Bounding Heuristic}

Simulation results using MCTS with and without the probabilistic bounding heuristic in Table \ref{tab:low_bounding} were tested for a process noise variation of 0.01, unknown parameter lower bound of 0.1, and confidence level of 0.95. The percentages denote the ratio of steps outside the desired bounds to total steps. Since these desired bounds are chosen by the control designer, several different bound levels were compared. The bounding heuristic decreased the percent out of bounds by a factor between 3 and 5. As expected with the conservative confidence region estimates, the state remains within the specified region more than 95\% of the time when the bounding heuristic is used for two of the three desired bounds. For very tight bounds, where there are few possible actions to keep the system within the desired bounds, the bounding heuristic approaches the same percentage out of the desired bounds as standard MCTS. The total reward values decreased with smaller desired bounds due to the extra constraint of only selecting actions that would meet the required next state norm. 

\begin{table}[!t]
\caption{Probabilistic Bounding Performance}
\label{tab:low_bounding}
\begin{center}
\resizebox{0.5\textwidth}{!}{
\begin{tabular}{cccccc} 
  \toprule \toprule
  Desired State Bounds & 6.0 & 5.0 & 4.0\\ 
  \midrule
  MCTS (percent outside bounds) & 2.5\% & 6.3\%&  27.3\%\\ 
  MCTS Heuristic (percent outside bounds) & 0.5\% & 2.2\% & 11.3\% \\
  Rewards with Heuristic & $-147.0$ & $-153.3$ & $-181.6$\\
 \bottomrule \bottomrule
\end{tabular}
}
\end{center}
\end{table}

\section{Conclusions}

This paper considers the problem of controlling a robot while estimating unknown parameters, a common challenge of physical systems in uncertain environments.
An online, sampling-based approach, MCTS, provides an approximate solution to this continuous control problem.
Simulations of a 2D manipulation task show that this method effectively balances exploration and exploitation, improving performance as the lower bound on the unknown parameters decreases.
This outperforms certainty equivalent MPC and QMDP-TS, an MCTS variant using a one-step lookahead to account for limited uncertainty.
An offline optimization method for automatically selecting tuning parameters and a heuristic for selecting actions to probabilistically bound the next state are also demonstrated.
The new MCTS algorithm addresses the challenge of estimation and control for many real-world systems that are nonlinear, have input constraints, can be modeled with discrete time steps, and have time-varying unknown parameters or any combinations of these attributes.


\section*{Acknowledgment}
The authors thank Mac Schwager for discussions on background material and the probabilistic bounding heuristic as well as Vincent Chow, Tori Fujinami, and Ian de Vlaming for their work on the QMDP-TS implementation.

\ifCLASSOPTIONcaptionsoff
  \newpage
\fi

%
%
\bibliographystyle{IEEEtran}
\bibliography{bio}

\begin{thebibliography}{10}
\providecommand{\url}[1]{#1}
\csname url@samestyle\endcsname
\providecommand{\newblock}{\relax}
\providecommand{\bibinfo}[2]{#2}
\providecommand{\BIBentrySTDinterwordspacing}{\spaceskip=0pt\relax}
\providecommand{\BIBentryALTinterwordstretchfactor}{4}
\providecommand{\BIBentryALTinterwordspacing}{\spaceskip=\fontdimen2\font plus
\BIBentryALTinterwordstretchfactor\fontdimen3\font minus
  \fontdimen4\font\relax}
\providecommand{\BIBforeignlanguage}[2]{{%
\expandafter\ifx\csname l@#1\endcsname\relax
\typeout{** WARNING: IEEEtran.bst: No hyphenation pattern has been}%
\typeout{** loaded for the language `#1'. Using the pattern for}%
\typeout{** the default language instead.}%
\else
\language=\csname l@#1\endcsname
\fi
#2}}
\providecommand{\BIBdecl}{\relax}
\BIBdecl

\bibitem{lavalle2001randomized}
S.~M. LaValle and J.~J. Kuffner, ``Randomized kinodynamic planning,''
  \emph{International Journal of Robotics Research}, vol.~20, no.~5, pp.
  378--400, 2001.

\bibitem{janson2015fast}
L.~Janson, E.~Schmerling, A.~Clark, and M.~Pavone, ``Fast marching tree: A fast
  marching sampling-based method for optimal motion planning in many
  dimensions,'' \emph{International Journal of Robotics Research}, vol.~34,
  no.~7, pp. 883--921, 2015.

\bibitem{thrun2005probabilistic}
S.~Thrun, W.~Burgard, and D.~Fox, \emph{Probabilistic Robotics}.\hskip 1em plus
  0.5em minus 0.4em\relax MIT Press, 2005.

\bibitem{d2008future}
R.~D'Andrea and P.~Wurman, ``Future challenges of coordinating hundreds of
  autonomous vehicles in distribution facilities,'' in \emph{Technologies for
  Practical Robot Applications}, 2008, pp. 80--83.

\bibitem{beiker2012legal}
S.~A. Beiker, ``Legal aspects of autonomous driving,'' \emph{Santa Clara L.
  Rev.}, vol.~52, p. 1145, 2012.

\bibitem{sadigh2016gathering}
D.~Sadigh, S.~S. Sastry, S.~A. Seshia, and A.~Dragan, ``Information gathering
  actions over human internal state,'' in \emph{International Conference on
  Intelligent Robots and Systems (IROS)}, 2016.

\bibitem{sunberg2017internal}
Z.~N. Sunberg, C.~J. Ho, and M.~J. Kochenderfer, ``The value of inferring the
  internal state of traffic participants for autonomous freeway driving,'' in
  \emph{American Control Conference (ACC)}.\hskip 1em plus 0.5em minus
  0.4em\relax IEEE, 2017.

\bibitem{pineau2003towards}
J.~Pineau, M.~Montemerlo, M.~Pollack, N.~Roy, and S.~Thrun, ``Towards robotic
  assistants in nursing homes: {Challenge} and results,'' \emph{Robotics and
  Autonomous Systems}, vol.~42, no.~3, pp. 271--281, 2003.

\bibitem{feldbaum1960dual}
A.~Feldbaum, ``Dual control theory. {I},'' \emph{Avtomatika i Telemekhanika},
  vol.~21, no.~9, pp. 1240--1249, 1960.

\bibitem{filatov2000survey}
N.~M. Filatov and H.~Unbehauen, ``Survey of adaptive dual control methods,''
  \emph{Control Theory and Applications}, vol. 147, no.~1, pp. 118--128, 2000.

\bibitem{ioannou1996robust}
P.~A. Ioannou and J.~Sun, \emph{Robust Adaptive Control}.\hskip 1em plus 0.5em
  minus 0.4em\relax PTR Prentice-Hall Upper Saddle River, NJ, 1996, vol.~1.

\bibitem{narendra2012stable}
K.~S. Narendra and A.~M. Annaswamy, \emph{Stable Adaptive Systems}.\hskip 1em
  plus 0.5em minus 0.4em\relax Courier Corporation, 2012.

\bibitem{bertsekas1995dynamic}
D.~P. Bertsekas, \emph{Dynamic Programming and Optimal Control}.\hskip 1em plus
  0.5em minus 0.4em\relax Athena Scientific, 1995, vol.~1, no.~2.

\bibitem{garcia1989model}
C.~E. Garcia, D.~M. Prett, and M.~Morari, ``Model predictive control: Theory
  and practice-—a survey,'' \emph{Automatica}, vol.~25, no.~3, pp. 335--348,
  1989.

\bibitem{chen1997quasi}
H.~Chen and F.~Allgower, ``A quasi-infinite horizon nonlinear model predictive
  control scheme with guaranteed stability,'' in \emph{European Control
  Conference (ECC)}.\hskip 1em plus 0.5em minus 0.4em\relax IEEE, 1997, pp.
  1421--1426.

\bibitem{liu2014nonlinear}
C.~Liu, A.~Gray, C.~Lee, J.~K. Hedrick, and J.~Pan, ``Nonlinear stochastic
  predictive control with unscented transformation for semi-autonomous
  vehicles,'' in \emph{American Control Conference (ACC)}.\hskip 1em plus 0.5em
  minus 0.4em\relax IEEE, 2014, pp. 5574--5579.

\bibitem{campo1987robust}
P.~J. Campo and M.~Morari, ``Robust model predictive control,'' in
  \emph{American Control Conference (ACC)}.\hskip 1em plus 0.5em minus
  0.4em\relax IEEE, 1987, pp. 1021--1026.

\bibitem{allwright1992linear}
J.~Allwright and G.~Papavasiliou, ``On linear programming and robust
  model-predictive control using impulse-responses,'' \emph{Systems \& Control
  Letters}, vol.~18, no.~2, pp. 159--164, 1992.

\bibitem{kouvaritakis2015model}
B.~Kouvaritakis and M.~Cannon, \emph{Model Predictive Control: {Classical},
  Robust and Stochastic}.\hskip 1em plus 0.5em minus 0.4em\relax Springer,
  2015.

\bibitem{mesbah2017stochastic}
A.~Mesbah, ``Stochastic model predictive control with active uncertainty
  learning: A survey on dual control,'' \emph{Annual Reviews in Control}, 2017.

\bibitem{streif2014stochastic}
S.~Streif, M.~Karl, and A.~Mesbah, ``Stochastic nonlinear model predictive
  control with efficient sample approximation of chance constraints,''
  \emph{arXiv preprint arXiv:1410.4535}, 2014.

\bibitem{bradford2017stochastic}
E.~Bradford and L.~Imsland, ``Stochastic nonlinear model predictive control
  with state estimation by incorporation of the unscented {Kalman} filter,''
  \emph{arXiv preprint arXiv:1709.01201}, 2017.

\bibitem{bavdekar2016stochastic}
V.~A. Bavdekar and A.~Mesbah, ``Stochastic nonlinear model predictive control
  with joint chance constraints,'' vol.~49, no.~18, pp. 270--275, 2016.

\bibitem{feng1991stability}
G.~Feng, C.~Zhang, and M.~Palaniswami, ``Stability analysis of input
  constrained continuous time indirect adaptive control,'' \emph{Systems \&
  Control Letters}, vol.~17, no.~3, pp. 209--215, 1991.

\bibitem{Wiering2012}
M.~Wiering and M.~van Otterlo, Eds., \emph{Reinforcement Learning: State of the
  Art}.\hskip 1em plus 0.5em minus 0.4em\relax New York: Springer, 2012.

\bibitem{kochenderfer2015decision}
M.~J. Kochenderfer, \emph{Decision Making Under Uncertainty: Theory and
  Application}.\hskip 1em plus 0.5em minus 0.4em\relax MIT Press, 2015.

\bibitem{bai2013planning}
H.~Bai, D.~Hsu, and W.~S. Lee, ``Planning how to learn,'' in
  \emph{International Conference on Robotics and Automation (ICRA)}.\hskip 1em
  plus 0.5em minus 0.4em\relax IEEE, 2013, pp. 2853--2859.

\bibitem{guez2013scalable}
A.~Guez, D.~Silver, and P.~Dayan, ``Scalable and efficient {Bayes}-adaptive
  reinforcement learning based on {Monte}-{Carlo} tree search,'' \emph{Journal
  of Artificial Intelligence Research}, vol.~48, pp. 841--883, 2013.

\bibitem{chen2016pomdplite}
M.~Chen, E.~Frazzoli, D.~Hsu, and W.~S. Lee, ``{POMDP}-lite for robust robot
  planning under uncertainty,'' in \emph{International Conference on Robotics
  and Automation (ICRA)}.\hskip 1em plus 0.5em minus 0.4em\relax IEEE, 2016,
  pp. 5427--5433.

\bibitem{webb2014online}
D.~J. Webb, K.~L. Crandall, and J.~van~den Berg, ``Online parameter estimation
  via real-time replanning of continuous {Gaussian POMDPs},'' in
  \emph{International Conference on Robotics and Automation (ICRA)}.\hskip 1em
  plus 0.5em minus 0.4em\relax IEEE, 2014, pp. 5998--6005.

\bibitem{slade2017simultaneous}
P.~Slade, P.~Culbertson, Z.~Sunberg, and M.~Kochenderfer, ``Simultaneous active
  parameter estimation and control using sampling-based {Bayesian}
  reinforcement learning,'' in \emph{International Conference on Intelligent
  Robots and Systems (IROS)}.\hskip 1em plus 0.5em minus 0.4em\relax IEEE/RSJ,
  2017, pp. 804--810.

\bibitem{sunberg2017pomcpow}
Z.~Sunberg and M.~Kochenderfer, ``Online algorithms for pomdps with continuous
  state, action, and observation spaces,'' in \emph{International Conference on
  Automated Planning and Scheduling ({ICAPS})}.

\bibitem{couetoux2011continuous}
A.~Cou{\"e}toux, J.~B. Hoock, N.~Sokolovska, O.~Teytaud, and N.~Bonnard,
  ``Continuous upper confidence trees,'' in \emph{International Conference on
  Learning and Intelligent Optimization}, 2011, pp. 433--445.

\bibitem{deng2006cross}
L.-Y. Deng, \emph{The Cross-Entropy Method: A Unified Approach to Combinatorial
  Optimization, {Monte-Carlo} Simulation, and Machine Learning}.\hskip 1em plus
  0.5em minus 0.4em\relax Taylor \& Francis, 2006.

\bibitem{powell2011}
W.~B. Powell, \emph{Approximate Dynamic Programming: Solving the Curses of
  Dimensionality}, 2nd~ed.\hskip 1em plus 0.5em minus 0.4em\relax Wiley, 2011.

\bibitem{papadimitriou1987complexity}
C.~H. Papadimitriou and J.~N. Tsitsiklis, ``The complexity of {Markov} decision
  processes,'' \emph{Mathematics of Operations Research}, vol.~12, no.~3, pp.
  441--450, 1987.

\bibitem{kaelbling1998planning}
L.~P. Kaelbling, M.~L. Littman, and A.~R. Cassandra, ``Planning and acting in
  partially observable stochastic domains,'' \emph{Artificial Intelligence},
  vol. 101, no.~1, pp. 99--134, 1998.

\bibitem{araya2010pomdp}
M.~Araya, O.~Buffet, V.~Thomas, and F.~Charpillet, ``A {POMDP} extension with
  belief-dependent rewards,'' in \emph{Advances in Neural Information
  Processing Systems}, 2010, pp. 64--72.

\bibitem{browne2012survey}
C.~B. Browne, E.~Powley, D.~Whitehouse, S.~M. Lucas, P.~I. Cowling,
  P.~Rohlfshagen, S.~Tavener, D.~Perez, S.~Samothrakis, and S.~Colton, ``A
  survey of {Monte Carlo} tree search methods,'' \emph{Transactions on
  Computational Intelligence and AI in Games}, vol.~4, no.~1, pp. 1--43, 2012.

\bibitem{bertsimas2014comparison}
D.~Bertsimas, J.~D. Griffith, V.~Gupta, M.~J. Kochenderfer, V.~V.
  Mi{\v{s}}i{\'c}, and R.~Moss, ``A comparison of {Monte Carlo} tree search and
  mathematical optimization for large scale dynamic resource allocation,''
  \emph{arXiv preprint arXiv:1405.5498}, 2014.

\bibitem{littman1995complexity}
M.~L. Littman, T.~L. Dean, and L.~P. Kaelbling, ``On the complexity of solving
  {Markov} decision problems,'' in \emph{Uncertainty in Artificial
  Intelligence}.\hskip 1em plus 0.5em minus 0.4em\relax Morgan Kaufmann
  Publishers Inc., 1995, pp. 394--402.

\bibitem{wan2000unscented}
E.~A. Wan and R.~Van Der~Merwe, ``The unscented kalman filter for nonlinear
  estimation,'' in \emph{Adaptive Systems for Signal Processing,
  Communications, and Control Symposium}.\hskip 1em plus 0.5em minus
  0.4em\relax IEEE, 2000, pp. 153--158.

\bibitem{julier1997new}
S.~J. Julier and J.~K. Uhlmann, ``New extension of the kalman filter to
  nonlinear systems,'' in \emph{Signal Processing, Sensor Fusion, and Target
  Recognition VI}, vol. 3068.\hskip 1em plus 0.5em minus 0.4em\relax
  International Society for Optics and Photonics, 1997, pp. 182--194.

\bibitem{beale1960confidence}
E.~Beale, ``Confidence regions in non-linear estimation,'' \emph{Journal of the
  Royal Statistical Society}, pp. 41--88, 1960.

\bibitem{draper2014applied}
N.~R. Draper and H.~Smith, \emph{Applied Regression Analysis}.\hskip 1em plus
  0.5em minus 0.4em\relax John Wiley \& Sons, 2014, vol. 326.

\bibitem{stoye2009more}
J.~Stoye, ``More on confidence intervals for partially identified parameters,''
  \emph{Econometrica}, vol.~77, no.~4, pp. 1299--1315, 2009.

\bibitem{cui2003empirical}
H.~Cui and S.~X. Chen, ``Empirical likelihood confidence region for parameter
  in the errors-in-variables models,'' \emph{Journal of Multivariate Analysis},
  vol.~84, no.~1, pp. 101--115, 2003.

\bibitem{egorov2017pomdps}
M.~Egorov, Z.~Sunberg, E.~Balaban, T.~Wheeler, J.~Gupta, and M.~Kochenderfer,
  ``{POMDPs.jl}: A framework for sequential decision making under
  uncertainty,'' \emph{Journal of Machine Learning Research}, 2017.

\bibitem{udell2014convex}
M.~Udell, K.~Mohan, D.~Zeng, J.~Hong, S.~Diamond, and S.~Boyd, ``Convex
  optimization in {Julia},'' in \emph{Workshop for High Performance Technical
  Computing in Dynamic Languages}.\hskip 1em plus 0.5em minus 0.4em\relax IEEE,
  2014, pp. 18--28.

\end{thebibliography}
%
%
%
%
%
%
\begin{IEEEbiography}[{\includegraphics[width=1in,height=1.25in,clip,keepaspectratio]{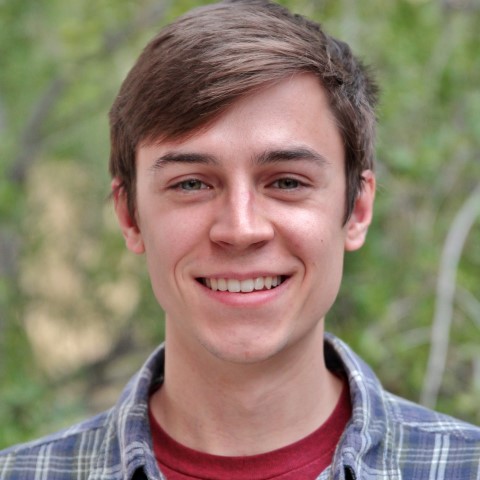}}]{Patrick Slade}
received the B.S. degree in mechanical engineering from the University of Illinois at Urbana-Champaign in 2016 and a M.S. in mechanical engineering from Stanford University in 2017. He is currently a Ph.D. candidate in mechanical engineering at Stanford University. His research focuses on intelligent and robust decision making for human-robot interaction with application to assistive devices. 
\end{IEEEbiography}
%
\begin{IEEEbiography}
[{\includegraphics[width=1in,height=1.25in,clip,keepaspectratio]{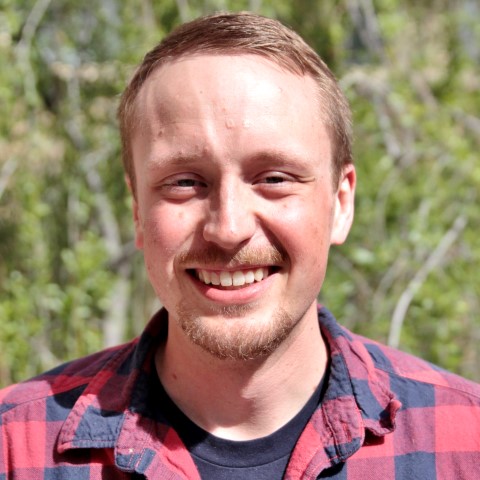}}]{Zachary N. Sunberg}
received the B.S. and M.S. degrees in aerospace engineering from Texas A\&M University in 2011 and 2013, respectively. He is currently a Ph.D. candidate in aerospace engineering at Stanford University. His research focuses on applying artificial intelligence to control autonomous vehicles in the physical world.
\end{IEEEbiography}
\begin{IEEEbiography}[{\includegraphics[width=1in,height=1.25in,clip,keepaspectratio]{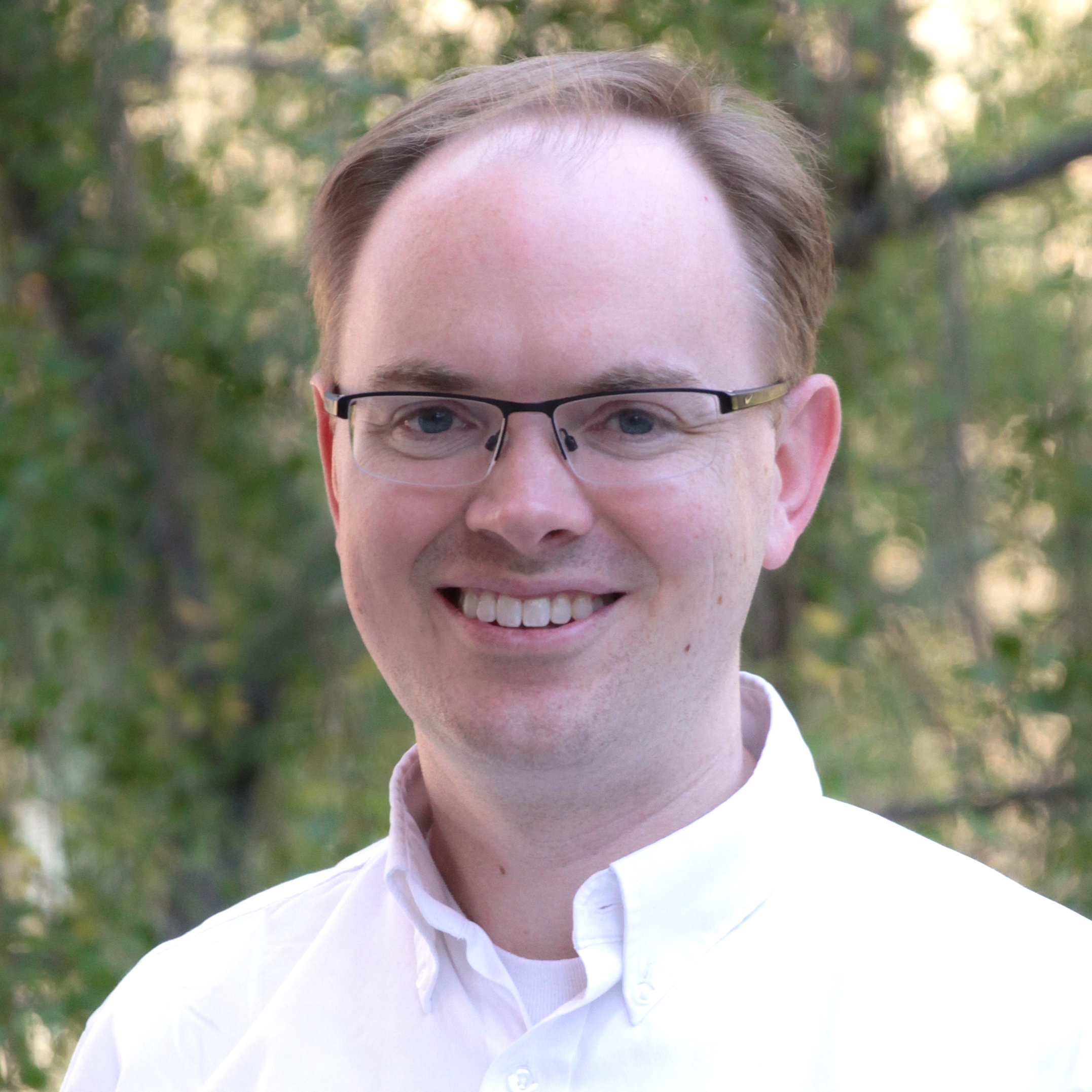}}]{Mykel J. Kochenderfer}
received the B.S. and M.S. degrees in computer science from Stanford University in 2003 and a Ph.D. studying at the Institute of Perception, Action, and Behaviour from the University of Edinburgh in 2006. He then joined MIT Lincoln Laboratory as technical staff for airspace modeling and aircraft collision avoidance. In 2013 he joined Stanford University as an Assistant Professor in the Aeronautics and Astronautics Department with a courtesy appointment in Computer Science. His research focuses on advanced algorithms and analytical methods for the design of robust decision making systems.
\end{IEEEbiography}


\vfill

\end{document}